%% file: paper_rev3.tex
\documentclass[12pt]{article}
\usepackage[T1]{fontenc}
\usepackage[latin9]{inputenc}
\usepackage{geometry}
\usepackage{verbatim}
\usepackage{float}
\usepackage{algorithm2e}
\usepackage{amsmath}
\usepackage{amsthm}
\usepackage{amssymb}
\usepackage{esint}
\usepackage[authoryear]{natbib}
\usepackage[unicode=true,
 bookmarks=false,
 breaklinks=true,pdfborder={0 0 0},pdfborderstyle={},backref=section,colorlinks=false]
 {hyperref}

\makeatletter
\theoremstyle{definition}
\newtheorem{defn}{\protect\definitionname}
\theoremstyle{remark}
\newtheorem{rem}{\protect\remarkname}
\theoremstyle{plain}

\theoremstyle{plain}

\advance \topmargin by -\headheight
\advance \topmargin by -\headsep

\usepackage[OT1]{fontenc}
\usepackage{amsthm,amsmath,amsfonts}
\usepackage{natbib}
\usepackage{booktabs}
\usepackage{mathrsfs}
\usepackage{graphicx,xcolor}
\usepackage{textcomp}
\usepackage[final]{microtype}
\usepackage{titling}

\LinesNumbered
\RestyleAlgo{algoruled} 

\providecommand{\propname}{Proposition}
\theoremstyle{plain}

\providecommand{\assumpname}{Assumption}
\theoremstyle{plain}

\makeatother

\providecommand{\definitionname}{Definition}
\providecommand{\lemmaname}{Lemma}
\providecommand{\remarkname}{Remark}
\providecommand{\theoremname}{Theorem}

\begin{document}
\input{macros.tex}
\title{A Unified Framework for Regularized Estimating Equations via Fixed-Point and Variational Inequality Problems}
\author{Archer Y. Yang \thanks{Co-first author; Co-corresponding author; Department of Mathematics and Statistics, McGill University; Mila (archer.yang@mcgill.ca)},
Yue Zhao \thanks{Co-first author; Department of Mathematics, University of York},
Yi Lian \thanks{Department of Biostatistics, Epidemiology and Informatics, University of Pennsylvania},
Yuwen Gu \thanks{Department of Statistics, University of Connecticut},
Jun Fan\thanks{Co-corresponding author, Department of Mathematics and Statistics, McGill
University}}
\maketitle
\begin{abstract}
Many statistics problems are formulated within an estimating equation framework instead of a minimization framework.  However, the regularized estimating equations (REE) have been much less extensively studies than regularized minimization problems.  In this paper, we study an improved regularized estimating equation formulation and explore its subsequent equivalences in terms of (1) fixed-point problem specified via the proximal operator of the corresponding regularizer, and (2) generalized variational inequality problems.  Such equivalences hold under general conditions and accommodate nonconvex regularizers.  Moreover, these equivalences open up new possibilities in theoretical analysis and computational algorithms when studying the REE.

\noindent %
\end{abstract}

\section{Introduction\label{sec:methodology}}

Suppose $\mathbf{U}(\beta)=(U_{1}(\beta),\ldots,U_{p}(\beta))^{\T}$
is an estimating function for $\beta=(\beta_{1},\ldots,\beta_{p})^{\T}$
based on a random sample of size $n$, where $\mathbf{U}(\cdot):\R^{p}\rightarrow\R^{p}$
is a vector-valued function. For example, in maximum likelihood estimation,
$\mathbf{U}(\beta)$ is the negative score function. In general, $\mathbf{U}(\beta)$
may not necessarily correspond to the negative gradient of an objective
function, such as a likelihood. Consider the standard estimating equation
\begin{equation}
\mathbf{U}(\beta)=\mathbf{0}.\label{eq:nonregularized_ee}
\end{equation}
Assume that the solution of \eqref{eq:nonregularized_ee} exists,
which is denoted by $\hat{\beta}$. Note that for any $\tau>0$,
\[
\mathbf{U}(\hat{\beta})=\mathbf{0}\qquad\Longleftrightarrow\quad\hat{\beta}=\hat{\beta}-\tau\mathbf{U}(\hat{\beta}).
\]
This motivates us to rewrite \eqref{eq:nonregularized_ee} as a \emph{fixed-point
problem:}
\begin{equation}
\text{find}\quad\hat{\beta}\in\R^{p}\quad\text{such that\ }\hat{\beta}=f(\hat{\beta}),\ \ \text{with }f(\beta)\equiv\beta-\tau\mathbf{U}(\beta).\label{eq:nonregul_fixed_pt}
\end{equation}
We can also show that \eqref{eq:nonregularized_ee} is equivalent
to the following \emph{variational inequality problem}:
\begin{equation}
\text{find}\quad\hat{\beta}\in\R^{p}\quad\text{such that\ensuremath{\quad}}\mathbf{U}(\hat{\beta})^{\top}(\beta-\hat{\beta})\geq0,\quad\text{for all }\beta\in\R^{p}.\label{eq:variational_unpenalized}
\end{equation}
This is because if $\mathbf{U}(\hat{\beta})=\mathbf{0}$, then inequality
\eqref{eq:variational_unpenalized} holds with equality for all $\beta$.
Conversely, if $\hat{\beta}$ satisfies \eqref{eq:variational_unpenalized},
we can choose $\beta=\hat{\beta}-\mathbf{U}(\hat{\beta})$, which
implies that $-\mathbf{U}(\hat{\beta})^{\top}\mathbf{U}(\hat{\beta})\geq0$
and therefore $\mathbf{U}(\hat{\beta})=\mathbf{0}$.

These results may have very little practical relevance, but it raises
an interesting question, that is, whether the equivalences between
estimating equations, fixed-point problems and variational inequality problems carry over to the regularization setting?

\section{Regularized estimating equations}

In this section, we extend the results to the more interesting regularization
cases. Existing literature on regularized estimating equations \citep{fu2003penalized,johnson2008penalized}
typically considers the following formulation: 
\begin{equation}
\mathbf{U}(\beta)+\mathbf{q}_{\lambda}(|\beta|)\odot\sgn(\beta)=\mathbf{0},\label{eq:old_ee}
\end{equation}
where $\sgn(\beta)=(\sgn(\beta_{1})\ddd\sgn(\beta_{p}))^{\T}$ and
$\mathbf{q}_{\lambda}(|\beta|)=(q_{\lambda}(|\beta_{1}|),\ldots,q_{\lambda}(|\beta_{p}|))^{\T}$
with $q_{\lambda}(\cdot)$ being a continuous function. Here $\odot$
denotes the component-wise product. The tuning parameter $\lambda>0$
determines the amount of regularization. \citet{johnson2008penalized}
mainly considered the case where $q_{\lambda}(|\beta_{j}|)=\frac{dp_{\lambda}(t)}{dt}\big|_{t=|\beta_{j}|}\equiv p_{\lambda}'(|\beta_{j}|)$
is the derivative of some penalty function $p_{\lambda}(\cdot)$ evaluated
at $|\beta_{j}|$ for $j=1\ddd p$. Some example penalties include
(a) the lasso penalty \citep{tibshirani96_regres_shrink_select_via_lasso},
$p_{\lambda}(|t|)=\lambda|t|$ ; (b) the elastic net penalty \citep{zou05_regul_variab_selec_via_elast_net},
$p_{\lambda}(|t|)=\lambda_{1}|t|+\lambda_{2}|t|^{2}$; and (c) the
SCAD penalty \citep{fan2001variable}, with derivative given by 
\[
p_{\lambda}^{\textup{scad},\prime}(|t|;a)=\lambda\left\{ I(|t|<\lambda)+\frac{(a\lambda-|t|)_{+}}{(a-1)\lambda}I(|t|\geq\lambda)\right\} ,
\]
for $a>2$ (see \eqref{eq:scad_mcp_penalty}). 

Note that formulation \eqref{eq:old_ee} only works for penalties
with element-wise separability and cannot be directly applied to many
other commonly-used penalties, such as the group lasso \citep{yuan2006model}
and the sparse group lasso \citep{simon2013sparse}. In this article,
we consider the regularized estimating equation in a slightly more
general form:
\begin{equation}
\mathbf{0}\in\mathbf{U}(\beta)+\lambda\partial\Omega(\beta),\label{eq:ee}
\end{equation}
where $\Omega(\cdot):\R^{p}\rightarrow\R$ is a general convex penalty
and the subdifferential $\partial\Omega(\beta)$ denotes the set of all subgradients of
$\Omega(\cdot)$ at $\beta$. A subgradient of $\Omega(\cdot)$ at
$\beta\in\R^{p}$ is defined as any vector $g\in\R^{p}$ such that
\[
\Omega(\beta')\geq\Omega(\beta)+g^{\T}(\beta'-\beta)\qquad\text{for all }\beta'.
\]
Note that $\partial\Omega(\beta)$ is a closed and convex set. Several
examples of formulation \eqref{eq:ee} follow.

\paragraph*{Ridge. }

If $\Omega(\cdot)$ is a convex and differentiable function, then
$\partial\Omega(\beta)=\{\nabla\Omega(\beta)\}$, i.e., the gradient
of $\Omega(\beta)$ at $\beta$ is its only subgradient. Therefore,
for the ridge penalty $\Omega(\beta)=\|\beta\|_{2}^{2}$, the sub-differential
set contains only the regular gradient $\partial\Omega(\beta)/\partial\beta=2\beta$
and thus \eqref{eq:ee} reduces to the regular estimating equation
$\mathbf{U}(\beta)+2\lambda\beta=\mathbf{0}$. 

\paragraph*{Lasso.}

If $\Omega(\cdot)$ is the lasso penalty $\Omega(\beta)=\|\beta\|_{1}$,
then $\beta$ must satisfy the equation
\begin{equation}
\mathbf{U}(\beta)+\lambda v=\mathbf{0},\label{eq:equivalent}
\end{equation}
where $v\in\partial\|\beta\|_{1}$ is a subgradient of $\|\beta\|_{1}$,
evaluated at $\beta$. The $j$-th element of $v$ is 
\begin{equation}
v_{j}=\begin{cases}
\sgn(\beta_{j}), & \text{if\ }\beta_{j}\neq0,\\
\in[-1,1], & \text{if\ }\beta_{j}=0,
\end{cases}\label{eq:subgrad_lasso}
\end{equation}
for $j=1\ddd p$. The estimating equation \eqref{eq:equivalent} yield
the following equivalent conditions
\begin{equation}
\begin{cases}
U_{j}(\beta)+\lambda\sgn(\beta_{j})=0, & \text{if\ \ }\beta_{j}\neq0,\\
|U_{j}(\beta)|\leq\lambda, & \text{if\ \ }\ensuremath{\beta_{j}}=0.
\end{cases}\label{eq:KKT_cond}
\end{equation}
Note that the first condition in \eqref{eq:KKT_cond} for $\beta_{j}\neq0$
coincides with the original formulation \eqref{eq:old_ee} by \citet{johnson2008penalized}
with $q_{\lambda}(|\beta_{j}|)=\lambda$, but \eqref{eq:old_ee} did
not explicitly handle the scenario $\ensuremath{\beta_{j}}=0$. When
$\mathbf{U}(\beta)=-X^{\top}(y-X\beta)$ is the negative gradient
of the least squares objective $L(\beta)=\frac{1}{2}\|y-X\beta\|^{2}$,
\eqref{eq:KKT_cond} corresponds to the KKT conditions of the lasso
regularized least squares problem. 

\paragraph*{Group lasso.}

Suppose the $p$ predictors are divided into several non-overlapping
groups. Let $\mathcal{G}=\{g_{1}\ddd g_{|\mathcal{G}|}\}$ be a partition
of the index set $\{1\ddd p\}$ into $|\mathcal{G}|$ groups. Each group $g_{j}$ is a subset of
the index set $\{1\ddd p\}$, with no overlaps with other groups,
that is $g_{j}\cap g_{k}=\varnothing$ for $k\neq j$.  The union of all $|\mathcal{G}|$ groups covers the entire index set
such that $\cup_{j=1}^{|\mathcal{G}|}g_{j}=\mathcal{G}$.  For the coefficient
vector $\beta=(\beta_{1}\ddd\beta_{p})^{\T}$, we let $\beta_{g}$
denote the sub-vector of $\beta$ whose indices are within $g\in\mathcal{G}$, and let $m_{g}$ be the size of group $g$. \citet{yuan2006model}
proposed the group lasso regularization $\Omega(\beta)=\sum_{g\in\mathcal{G}}\sqrt{m_{g}}\|\beta_{g}\|_{2}$.
For ease of notation, we omit the weights $\sqrt{m_{g}}$ in the penalty
term. The corresponding regularized estimating equation is 
\begin{equation}
\mathbf{0}\in\mathbf{U}(\beta)+\lambda\partial\left(\sum_{g\in\mathcal{G}}\|\beta_{g}\|_{2}\right).\label{eq:group_ee}
\end{equation}
Denote by $[x]_{g}$ the sub-vector of $x$ for group $g$. The solution
to \eqref{eq:group_ee} satisfies the following equation, for each
group $g$:
\[
[\mathbf{U}(\beta)]_{g}+\lambda u_{g}=\mathbf{0},
\]
where $u_{g}$ is the subgradient of $\|\beta_{g}\|_{2}$ evaluated
at $\beta_{g}$ with
\begin{equation}
u_{g}=\begin{cases}
\frac{\beta_{g}}{\|\beta_{g}\|_{2}}, & \text{if\ }\beta_{g}\neq\mathbf{0},\\
\in\{x:\|x\|_{2}\leq1\}, & \text{if\ }\beta_{g}=\mathbf{0}.
\end{cases}\label{eq:subgrad_group_lasso}
\end{equation}
The subgradient equation \eqref{eq:group_ee} yields the following
equivalent conditions
\[
\begin{cases}
[\mathbf{U}(\beta)]_{g}+\lambda\frac{\beta_{g}}{\|\beta_{g}\|_{2}}=\mathbf{0}, & \text{if }\beta_{g}\neq\mathbf{0},\\
\|[\mathbf{U}(\beta)]_{g}\|_{2}\leq\lambda, & \text{if }\beta_{g}=\mathbf{0}.
\end{cases}
\]

\paragraph*{Sparse group lasso.}

As an important extension of the group lasso, \citet{simon2013sparse}
proposed the sparse group lasso which allows both group-wise and within-group
sparsity. The penalty is a convex combination of the lasso and group-lasso
penalties, $\Omega(\beta)=\sum_{g\in\mathcal{G}}(1-\alpha)\|\beta_{g}\|_{2}+\alpha\|\beta\|_{1}$,
where $\alpha\in[0,1]$. For each group $g$, the corresponding regularized
estimating equation is 
\begin{equation}
[\mathbf{U}(\beta)]_{g}+\lambda(1-\alpha)u_{g}+\lambda\alpha v_{g}=\mathbf{0},\label{eq:sparse_group_ee}
\end{equation}
where $u_{g}$ is a subgradient as defined in \eqref{eq:subgrad_group_lasso}
and $v_{g}$ is the sub-vector of a subgradient $v$ as defined in
\eqref{eq:subgrad_lasso}.

\section{Fixed-point formulation}
\label{sec:connect_REE_FPP}

In this section, we provide a connection between regularized estimating
equations and fixed-point problems. Assume that the solution of \eqref{eq:ee}
exists, which we denote by $\hat{\beta}$. Then we have the following
equivalent conditions for $\tau>0$:
\begin{eqnarray}
 & \mathbf{0} & \in\mathbf{U}(\hat{\beta})+\lambda\partial\Omega(\hat{\beta})\nonumber \\
\Longleftrightarrow\quad & \mathbf{0} & \in\hat{\beta}-(\hat{\beta}-\tau\mathbf{U}(\hat{\beta}))+\tau\lambda\partial\Omega(\hat{\beta})\label{eq:optim:cond}\\
\Longleftrightarrow\quad & \mathbf{0} & \in\frac{1}{2}\nabla_{\beta}\|\beta-(\hat{\beta}-\tau\mathbf{U}(\hat{\beta}))\|_{2}^{2}\Big|_{\beta=\hat{\beta}}+\tau\lambda\partial_{\beta}\Omega(z)\Big|_{\beta=\hat{\beta}},\nonumber 
\end{eqnarray}
where the differentiation $\nabla_{\beta}$ and subdifferential $\partial_{\beta}$
are with respect to $\beta$. If $\Omega(\beta)$ is a convex penalty,
the last line of \eqref{eq:optim:cond} characterizes the necessary
and sufficient condition for $\hat{\beta}$ to be a minimizer of the
penalized quadratic function:
\begin{align}
\hat{\beta}=\argmin_{\beta}\frac{1}{2}\|\beta-(\hat{\beta}-\tau\mathbf{U}(\hat{\beta}))\|_{2}^{2}+\tau\lambda\Omega(\beta).\label{eq:minimization}
\end{align}
Let $\mathrm{prox}_{\Omega}:\R^{p}\rightarrow\R^{p}$ denote the \emph{proximal
operator }\citep{parikh2014proximal} of the convex penalty function
$\Omega$, 
\begin{equation}
\mathrm{prox}_{\Omega}(v)=\argmin_{z}\frac{1}{2}\|z-v\|_{2}^{2}+\Omega(z).\label{eq:subprob_prox}
\end{equation}
Since the regularized quadratic function on the right-hand side of
\eqref{eq:subprob_prox} is strongly convex, it has a unique minimizer
for every $v\in\R^{p}$. Now we can rewrite \eqref{eq:minimization}
as a fixed-point problem:
\begin{equation}
\text{find}\quad\hat{\beta}\in\R^{p}\quad\text{such that\ }\hat{\beta}=f(\hat{\beta}),\ \ \text{with }f(\beta)\equiv\mathrm{prox}_{\tau\lambda\Omega}(\beta-\tau\mathbf{U}(\beta)).\label{eq:fixed_pt}
\end{equation}
Therefore $\hat{\beta}$ is a solution to \eqref{eq:ee} if and only
if $\hat{\beta}$ is a solution to \eqref{eq:fixed_pt}. Note that
if $\lambda=0$, the operator $\mathrm{prox}_{\tau\lambda\Omega}(v)$
reduces to $v$, thus \eqref{eq:fixed_pt} simplifies to \eqref{eq:nonregul_fixed_pt}.

Evaluating the proximal operator of a convex penalty $\Omega$ requires solving a small strongly convex optimization problem \eqref{eq:subprob_prox}.
In many cases, these problems often admit closed form solutions or can be solved very efficiently using specialized algorithms. We present several examples below.

\paragraph*{Lasso.}

When the penalty is lasso, the $j$-th element of the proximal operator
is
\begin{align*}
[\text{prox}_{\tau\lambda\|\cdot\|_{1}}(v)]_{j} & =\sgn(v_{j})(|v_{j}|-\tau\lambda)_{+}\equiv S_{\tau\lambda}(v_{j}),
\end{align*}
which is the soft-thresholding rule.

\paragraph*{Group lasso. }

The group lasso penalty has a closed form proximal operator \citep{parikh2014proximal}:
for group $g$,
\begin{align*}
[\text{prox}_{\tau\lambda\Omega}(v)]_{g} & =\biggl(1-\frac{\tau\lambda}{\|z_{g}\|_{2}}\biggr)_{+}v_{g},
\end{align*}
where $[x]_{g}$ is the sub-vector corresponding to group $g$ of
$x$.

\paragraph*{Sparse group lasso.}

The sparse group lasso also has a closed form proximal operator \citep{simon2013sparse}:
for group $g$,
\begin{align*}
[\text{prox}_{\tau\lambda\Omega}(v)]_{g} & =\argmin_{z_{g}}\frac{1}{2}\|z_{g}-v_{g}\|_{2}^{2}+\tau\lambda\left[(1-\alpha)\|z_{g}\|_{2}+\alpha\|z_{g}\|_{1}\right]\\
 & =\left[\left(1-\frac{(1-\alpha)\tau\lambda}{\|S_{\alpha\tau\lambda}(v_{g})\|_{2}}\right)_{+}S_{\alpha\tau\lambda}(v_{g})\right]_{g},
\end{align*}
where $S_{\alpha\tau\lambda}(v_{g})\equiv(S_{\alpha\tau\lambda}([v_{g}]_{1})\ddd S_{\alpha\tau\lambda}([v_{g}]_{m_{g}}))^{\top}$
with $S_{\alpha\tau\lambda}(x)=\sgn(x)(|x|-\alpha\tau\lambda)_{+}$.

\section{Variational inequality formulation}
\label{sec:REE_to_VIP}

After establishing the equivalences between regularized estimating
equations and fixed-point problems, we also show a connection between
regularized estimating equations and variational inequality problems.
This is not surprising since equivalences between fixed-point problems
and variational inequality problems are well known \citep[see,  e.g., ][]{malitsky2019golden}. 

Following \eqref{eq:ee}, a solution $\hat{\beta}$ should satisfy
$-\mathbf{U}(\hat{\beta})/\lambda\in\partial\Omega(\hat{\beta})$.
This implies that $-\mathbf{U}(\hat{\beta})/\lambda$ is a subgradient
of $\Omega$ at $\hat{\beta}$. Thus, by the definition of a subgradient,
\begin{equation}
\Omega(\beta)\geq\Omega(\hat{\beta})-\mathbf{U}(\hat{\beta})^{\top}(\beta-\hat{\beta})/\lambda\label{eq:subgradient_penalized}
\end{equation}
for any $\beta$. It follows that \eqref{eq:subgradient_penalized}
can be rewritten as a variational inequality problem:
\begin{equation}
\text{find}\quad\hat{\beta}\in\R^{p}\quad\text{such that\ensuremath{\quad}}\mathbf{U}(\hat{\beta})^{\top}(\beta-\hat{\beta})+\lambda(\Omega(\beta)-\Omega(\hat{\beta}))\geq0,\quad\text{for all }\beta\in\R^{p}.\label{eq:VIP}
\end{equation}
Note that if $\lambda=0$, \eqref{eq:VIP} reduces to \eqref{eq:variational_unpenalized}.
Unlike formulations \eqref{eq:ee} and \eqref{eq:fixed_pt}, which
require either specification of the subgradient or evaluation of the proximal operator of $\Omega$, formulation \eqref{eq:VIP} only
needs us to specify $\mathbf{U}(\beta)$ and $\Omega(\beta)$.

\section{Extensions to constrained forms}

Alternative to the Lagrangian form \eqref{eq:ee}, one may also consider
the constrained form of regularized estimating equations
\begin{equation}
\mathbf{U}(\beta)=\mathbf{0},\qquad\text{such that }\beta\in\mathcal{C},\label{eq:constrained_ee}
\end{equation}
where $\mathcal{C}$ is a convex set. For example, $\mathcal{C}$
can be a normed ball $\{\beta:\Phi(\beta)\leq r\}$ with the norm
function $\Phi(\cdot)$ and radius $r>0$. One can set $\Phi(\beta)$
to be $\|\beta\|_{1}$ for the lasso constraint, and $\sum_{g\in\mathcal{G}}\|\beta_{g}\|_{2}$
for the group lasso constraint, etc. Intriguingly, \eqref{eq:constrained_ee}
can still be viewed as an instance of \eqref{eq:ee}. Let $I_{\mathcal{C}}(\beta):\R^{p}\rightarrow\R$
be an indicator function
\[
I_{\mathcal{C}}(\beta)=\begin{cases}
0 & \text{if }\beta\in\mathcal{C}\\
\infty & \text{if }\beta\notin\mathcal{C}.
\end{cases}
\]
Assume the solution of \eqref{eq:constrained_ee} exists, then \eqref{eq:constrained_ee}
is equivalent to \eqref{eq:ee} with $\Omega(\beta)=I_{\mathcal{C}}(\beta)$
and $\lambda=1$. Let $\hat{\beta}$ be the solution of \eqref{eq:constrained_ee},
the fixed-point formulation thus apply
\begin{eqnarray}
 & \mathbf{0} & \in\mathbf{U}(\hat{\beta})+\partial I_{\mathcal{C}}(\hat{\beta})\nonumber \\
\Longleftrightarrow\quad & \hat{\beta} & =\mathrm{prox}_{\tau I_{\mathcal{C}}}(\hat{\beta}-\tau\mathbf{U}(\hat{\beta}))\nonumber \\
\Longleftrightarrow\quad & \hat{\beta} & =\argmin_{\beta\in\mathcal{C}}\frac{1}{2}\|\beta-(\hat{\beta}-\tau\mathbf{U}(\hat{\beta}))\|_{2}^{2}\nonumber \\
\Longleftrightarrow\quad & \hat{\beta} & =P_{\mathcal{C}}(\hat{\beta}-\tau\mathbf{U}(\hat{\beta})),\label{eq:projection}
\end{eqnarray}
where the projection operator onto $\mathcal{C}$ is defined as 
\[
P_{\mathcal{C}}(y)=\argmin_{x\in\mathcal{C}}\frac{1}{2}\|x-y\|_{2}^{2}.
\]
From \eqref{eq:projection} we can see that the proximal operator
associated with the constraint $I_{\mathcal{C}}(\hat{\beta})$ becomes
the projection onto the convex set $\mathcal{C}$, which shows that
\eqref{eq:constrained_ee} can be rewritten as the fixed-point problem
\[
\text{find}\quad\hat{\beta}\in\R^{p}\quad\text{such that\ }\hat{\beta}=f(\hat{\beta}),\ \ \text{with }f(\beta)\equiv P_{\mathcal{C}}(\beta-\tau\mathbf{U}(\beta)).
\]

On the other hand, \eqref{eq:constrained_ee} can also be represented
as the variational inequality problem 
\[
\text{find}\quad\hat{\beta}\in\R^{p}\quad\text{such that\ensuremath{\quad}}\mathbf{U}(\hat{\beta})^{\top}(\beta-\hat{\beta})\geq0,\quad\text{for all }\beta\in\mathcal{C}.
\]
To see this, let $\mathcal{N}_{\mathcal{C}}(\beta)$ be the normal
cone of $\mathcal{C}$ at $\beta$, 
\[
\mathcal{N}_{\mathcal{C}}(\beta)=\{g\in\R^{p}:g^{\top}(\beta'-\beta)\leq0\quad\text{for all }\beta'\in\mathcal{C}\}.
\]
For $\beta\in\mathcal{C}$, we know that $\partial I_{\mathcal{C}}(\beta)=\mathcal{N}_{\mathcal{C}}(\beta)$,
which gives
\begin{align*}
 & \mathbf{0}\in\mathbf{U}(\hat{\beta})+\partial I_{\mathcal{C}}(\hat{\beta})=\mathbf{U}(\hat{\beta})+\mathcal{N}_{\mathcal{C}}(\hat{\beta})\\
\Longleftrightarrow\quad & -\mathbf{U}(\hat{\beta})\in\mathcal{N}_{\mathcal{C}}(\hat{\beta})\\
\Longleftrightarrow\quad & \mathbf{U}(\hat{\beta})^{\top}(\beta-\hat{\beta})\geq0\qquad\text{for all }\beta\in\mathcal{C},
\end{align*}
as desired.

\section{Extensions to nonconvex penalties}

In this section, we extend the discussions in Sections~\ref{sec:connect_REE_FPP} and \ref{sec:REE_to_VIP} by establishing the connections of regularized estimating equations with \textit{nonconvex} penalties, which we often simply refer to in this section as nonconvex REEs, also to the formulations in terms of fixed-point problems (Section~\ref{sec:connect_REE_FPP_nonconvex}) and variational inequality problems (Section~\ref{sec:equiv_REE_VIP_nonconvex}).  Once the said connections are established, the corresponding algorithmic implementations will follow in a manner analogous to what we will show for the convex penalty cases in Sections~\ref{sec:connect_REE_FPP_algorithm} and \ref{sec:REE_to_VIP_algorithm}; for brevity, we omit the implementation details.

\subsection{The Clarke subdifferential, and new formulation of nonconvex REEs}

We use $\Omega_{\lambda}^{\nc}(\cdot):\RR^p\rightarrow\RR$ to denote a nonconvex penalty, such as the SCAD \citep{fan2001variable} or the MCP \citep{Zhang2010}, with regularization parameter $\lambda>0$.  Neither the SCAD nor the MCP penalty admits a (regular) subdifferential except at zero, and therefore the corresponding nonconvex REE cannot be directly defined as in \eqref{eq:ee}.  We instead adopt the \emph{Clarke subdifferential} \citep{clarke1990optimization} as a natural generalization of the notion of subdifferential that enables our analysis of nonconvex penalties. To start, we first recall the definition of the class of locally Lipschitz functions: 
\begin{defn}[Locally Lipschitz function]
\label{def:local_lips}
A function $f:S\rightarrow\R$ is locally Lipschitz at $x_0\in S$ if there exists a neighborhood $V$ of $x_0$ and a constant $L>0$ (both of which may depend on $x_0$) such that
\begin{align*}
    |f(x)-f(y)|\le L\|x-y\|_{2}\quad\text{ for all }x,y\in V\cap S.
\end{align*}
A function $f:S\rightarrow\R$ is locally Lipschitz on $S$ if $f$ is locally Lipschitz at every $x_0\in S$.
\end{defn} 
The classical Rademacher theorem (see for instance Theorem~9.60 in \cite{rockafellar2009variational}) states that a locally Lipschitz function $f$ is differentiable almost everywhere. In particular, every neighborhood of $x$ contains a point $y$ for which $\nabla f(y)$ exists. The Clarke subdifferential of a locally Lipschitz function $f$ at $x_{0}$, denoted by $\partial_{C}f(x_{0})$, can then be defined as (see for instance Theorem~2.5.1 in \cite{clarke1990optimization}), 
\begin{align*}
    \textstyle \partial_{C}f(x_{0})=\mathbf{Co}\left\{ \lim_{k\rightarrow\infty}\nabla f(x^{k})\ \big\lvert\ x^{k}\rightarrow x_{0}, f\text{ is differentiable at }x^{k}\right\},
\end{align*}
where $\mathbf{Co}$ denotes the convex hull of the argument set. Most commonly-used nonconvex penalties are locally Lipschitz, and hence admit Clarke subdifferentials everywhere on their domain. For a convex penalty, its Clarke subdifferential reduces to the regular subdifferential.

With the introduction of the Clarke subdifferential, we propose to write the nonconvex REE as 
\begin{equation}
\bfz\in\mathbf{U}(\beta)+\partial_{C}\Omega_{\lambda}^{\nc}(\beta), \label{eq:noncvx_REE}
\end{equation}
which differs from the convex REE~\eqref{eq:ee} only in the substitution by the Clarke subdifferential.

\subsection{Connection of nonconvex REEs to fixed-point problems}
\label{sec:connect_REE_FPP_nonconvex}

In general, nonconvex REEs may not give rise to fixed-point problems and variational inequality problems.  However, if the nonconvex penalty $\Omega_{\lambda}^{\nc}$ satisfies the following \emph{weak convexity} property, an updated equivalence between the nonconvex REEs, fixed-point problems and variational inequality problems can again be established.
\begin{defn}[Weak convexity]
\label{def:weak_convexity}
A function $\rho:\R^s\rightarrow\R$ (where $s$ is some generic dimension) is $\mu$-weakly convex if $\rho(v)+\frac{\mu}{2}\|v\|_2^{2}$ is convex, for some $\mu>0$.  Here $\mu$ controls the level of nonconvexity.  By Propositions~4.3 and 4.4 in \cite{Vial1983}, such a function $\rho$ is locally Lipschitz (and hence admits Clarke subdifferentials).
\end{defn}
\begin{rem}
The univariate SCAD penalty \citep{fan2001variable} and MCP penalty \citep{Zhang2010} with regularization parameter $\lambda>0$ are given respectively by
\begin{equation}
p_{\lambda}^{\textup{scad}}(t;a)=\begin{cases}
\lambda|t|, & \text{ for }|t|\le\lambda,\\
-\frac{t^{2}-2a\lambda|t|+\lambda^{2}}{2(a-1)}, & \text{ for }\lambda<|t|\le a\lambda,\\
\frac{(a+1)\lambda^{2}}{2}, & \text{ for }|t|>a\lambda,
\end{cases} \quad\quad p_{\lambda}^{\textup{mcp}}(t;b)=\begin{cases}
\lambda|t|-\frac{t^{2}}{2b}, & \text{ if }|t|\le b\lambda,\\
\frac{1}{2}b\lambda^{2}, & \text{ if }|t|>b\lambda,
\end{cases}
\label{eq:scad_mcp_penalty}
\end{equation}
where $a>2$ (for SCAD) and $b>0$ (for MCP) are penalty-specific, adjustable parameters.
The univariate SCAD and MCP penalties $p_{\lambda}^{\textup{scad}}(\cdot;a)$ and $p_{\lambda}^{\textup{mcp}}(\cdot;b)$ are $\mu$-weakly convex with $\mu=\frac{1}{a-1}$ and $\mu=\frac{1}{b}$, respectively; see for instance Appendix~A.1 in \cite{LohWainwright2017}.  Moreover, it is easily shown that the multivariate SCAD and MCP penalties $\Omega_{\lambda}^{\nc}(\beta)=\sum_{j=1}^{p}p_{\lambda}(\beta_{j})$, where $p_{\lambda}(\cdot)=p_{\lambda}^{\textup{scad}}(\cdot;a)$ or $p_{\lambda}(\cdot)=p_{\lambda}^{\textup{mcp}}(\cdot;b)$, are also $\mu$-weakly with the same $\mu$.
\end{rem}

Now we are ready to establish the equivalence between fixed-point problems on the one hand and nonconvex REEs involving a nonconvex penalty $\Omega_{\lambda}^{\nc}$ that is $\mu$-weakly convex on the other.  In terms of the Clarke subdifferential $\partial_{C}$ (henceforth always understood to be with respect to $\beta$), the following equivalences hold trivially for any $\tau>0$:
\begin{eqnarray*}
 & \bfz & \in\mathbf{U}(\hat\beta)+\partial_{C}\Omega_{\lambda}^{\nc}(\hat\beta)\\
\Longleftrightarrow\quad & \bfz & \in\hat\beta-(\hat\beta-\tau\mathbf{U}(\hat\beta))+\tau\partial_{C}\Omega_{\lambda}^{\nc}(\hat\beta)\\
\Longleftrightarrow\quad & \bfz & \textstyle \in\frac{1}{2}\partial_{C}\|\beta-(\hat\beta-\tau\mathbf{U}(\hat\beta))\|_{2}^{2}\Big|_{\beta=\hat\beta}+\tau\partial_{C}\Omega_{\lambda}^{\nc}(\beta)\Big|_{\beta=\hat\beta} .
\end{eqnarray*}
Now, both $\|\beta-(\hat\beta-\tau\mathbf{U}(\hat\beta))\|_{2}^{2}$
and $\Omega_{\lambda}^{\nc}(\beta)$ are locally Lipschitz, and so (by the sum rule in Fact~4 in \cite{LiSoMa2020})
\begin{align*}
\textstyle \frac{1}{2}\partial_{C}\|\beta-(\hat\beta-\tau\mathbf{U}(\hat\beta))\|_{2}^{2}+\tau\partial_{C}\Omega_{\lambda}^{\nc}(\beta)=\partial_{C}\left\{ \frac{1}{2}\|\beta-(\hat\beta-\tau\mathbf{U}(\hat\beta))\|_{2}^{2}+\tau\Omega_{\lambda}^{\nc}(\beta)\right\}.
\end{align*}
Because $\Omega_{\lambda}^{\nc}$ is assumed to be $\mu$-weakly convex, for a sufficiently small $\tau$ ($\tau<\frac{1}{\mu}$ suffices), the function inside the curly bracket on the right-hand side of the above equation is strongly convex in $\beta$.  Then, the Clarke subdifferential $\partial_{C}$ reduces to the (regular) subdifferential $\partial$, so 
\begin{align*}
\textstyle \bfz\in\partial\left\{ \frac{1}{2}\|\beta-(\hat\beta-\tau\mathbf{U}(\hat\beta))\|_{2}^{2}+\tau\Omega_{\lambda}^{\nc}(\beta)\right\} \Big|_{\beta=\hat\beta} .
\end{align*}
This characterizes the necessary and sufficient condition for $\hat\beta$ to be the minimizer of a convex objective: 
\begin{align*}
\textstyle \hat\beta=\argmin_{\beta}\frac{1}{2}\|\beta-(\hat\beta-\tau\mathbf{U}(\hat\beta))\|_{2}^{2}+\tau\Omega_{\lambda}^{\nc}(\beta).
\end{align*}
Thus the nonconvex REE \eqref{eq:noncvx_REE} can again be rewritten as a fixed-point problem
\begin{align*}
\text{find}\quad\hat\beta & \in\R^p\quad \text{such that}\quad\hat\beta=f(\hat\beta), \quad \text{with \ensuremath{\quad}} f(\beta)\equiv\mathrm{prox}_{\tau\Omega_{\lambda}^{\nc}}(\beta-\tau\mathbf{U}(\beta)).
\end{align*}

\begin{rem}
The proximal operators of some popular nonconvex penalties, such as the SCAD and the MCP, are also available in closed form and are easy to compute.
\end{rem}

\subsection{Connection of nonconvex REEs to variational inequality problems}
\label{sec:equiv_REE_VIP_nonconvex}

Now we show the connection between the nonconvex REEs (that involve a $\mu$-weakly convex $\Omega_{\lambda}^{\nc}$) and variational inequality problems. In terms of the Clarke subdifferential $\partial_{C}$, 
\begin{gather*}
\bfz \in\mathbf{U}(\hat\beta)+\partial_{C}\Omega_{\lambda}^{\nc}(\hat\beta) 
\Longleftrightarrow \bfz \textstyle \in\mathbf{U}(\hat\beta)-\mu\hat\beta+\partial_{C}\left\{ \Omega_{\lambda}^{\nc}(\beta)+\frac{\mu}{2}\|\beta\|_{2}^{2}\right\} \big|_{\beta=\hat\beta}\\
\Longleftrightarrow\quad -\mathbf{U}(\hat\beta)+\mu\hat\beta \textstyle \in\partial_{C}\left\{ \Omega_{\lambda}^{\nc}(\beta)+\frac{\mu}{2}\|\beta\|_{2}^{2}\right\} \big|_{\beta=\hat\beta} \Longleftrightarrow \textstyle -\mathbf{U}(\hat\beta)+\mu\hat\beta \textstyle  \in\partial\left\{ \Omega_{\lambda}^{\nc}(\beta)+\frac{\mu}{2}\|\beta\|_{2}^{2}\right\} \big|_{\beta=\hat\beta},
\end{gather*}
where the last step follows because $\Omega_{\lambda}^{\nc}(\beta)+\frac{\mu}{2}\|\beta\|_{2}^{2}$
is convex by the weak convexity of $\Omega_{\lambda}^{\nc}$.  Then, by the definition of a subgradient, instead of \eqref{eq:subgradient_penalized}, this time we have
\begin{align*}
    \textstyle \Omega_{\lambda}^{\nc}(\beta)+\frac{\mu}{2}\|\beta\|_{2}^{2}\ge\Omega_{\lambda}^{\nc}(\hat\beta)+\frac{\mu}{2}\|\hat\beta\|_{2}^{2}+(-\mathbf{U}(\hat\beta)+\mu\hat\beta) ^{\top}(\beta-\hat\beta)
\end{align*}
for any $\beta$. The above then leads to the following variational inequality problem that is the counterpart to \eqref{eq:VIP} now under a nonconvex penalty:
\begin{align*}
\textstyle \text{find}\ \hat\beta\in\R^p\ \text{such that}\ &( \mathbf{U}(\hat\beta)-\mu\hat\beta) ^{\top}(\beta-\hat\beta) \nonumber \\
&\textstyle  +\left\{ \Omega_{\lambda}^{\nc}(\beta) + \frac{\mu}{2}\|\beta\|_{2}^{2}-\Omega_{\lambda}^{\nc}(\hat\beta)-\frac{\mu}{2}\|\hat\beta\|_{2}^{2}\right\} \ge0,\ \text{for all}\ \beta\in\R^p. 
\end{align*}
Just as in the convex penalty case in Section~\ref{sec:REE_to_VIP}, the variational inequality formulation above does not require expressions of the Clarke subgradient or the proximal operator of $\Omega_{\lambda}^{\nc}$.

\section{Computation\label{sec:algorithm}}

Formulations \eqref{eq:fixed_pt} and \eqref{eq:VIP} reveal interesting
connections between regularized estimating equations, fixed-point
problems and variational inequality problems. To solve large-scale regularized
estimating equations, it might be worth pursuing computation from
\eqref{eq:fixed_pt} and \eqref{eq:VIP}. While fast computational
algorithms are less developed for \eqref{eq:old_ee}, there are many
efficient solvers for fixed-point problems and variational inequality problems.
In this regard, we apply some efficient and scalable solvers to \eqref{eq:fixed_pt}
and \eqref{eq:VIP}, and examine their performance against existing
algorithms for regularized estimating equations.

\subsection{Existing approaches}

To solve \eqref{eq:old_ee}, many existing works \citep[e.g.,][]{johnson2008penalized}
adopted the local quadratic approximation (LQA) approach proposed
by \citet{fan2001variable}. Specifically, they considered a local
quadratic approximation to the penalty function
\[
p_{\lambda}(|\beta_{j}|)\approx p_{\lambda}(|\tilde{\beta}_{j}|)+\frac{1}{2}\frac{p_{\lambda}'(|\tilde{\beta}_{j}|)}{|\tilde{\beta}_{j}|}(\beta_{j}^{2}-\tilde{\beta}_{j}^{2})
\]
around an iterate $\tilde{\beta}_{j}$. This yields the following
approximation to the subgradient of $p_{\lambda}(|\beta_{j}|)$ when
$\tilde{\beta}_{j}\neq0$:

\begin{equation}
\frac{\partial}{\partial\beta_{j}}p_{\lambda}(|\beta_{j}|)=p_{\lambda}^{\prime}(|\beta_{j}|)\sgn(\beta_{j})\approx\frac{p_{\lambda}^{\prime}(|\tilde{\beta}_{j}|)}{|\tilde{\beta}_{j}|}\beta_{j}.\label{eq:approx}
\end{equation}
By this local quadratic approximation, the Newton\textendash Raphson
algorithm was used to solve the following equation
\begin{equation}
Q_{\widetilde{\beta}}(\beta):=\mathbf{U}(\beta)+\boldsymbol{\Lambda}_{\lambda}(\widetilde{\beta})\odot\beta=\mathbf{0},\label{eq:LQA}
\end{equation}
where $\boldsymbol{\Lambda}_{\lambda}(\widetilde{\beta})=\diag\{p_{\lambda}^{\prime}(|\tilde{\beta}_{1}|)/|\tilde{\beta}_{1}|\ddd p_{\lambda}^{\prime}(|\tilde{\beta}_{p}|)/|\tilde{\beta}_{p}|\}$.
Let $\beta^{(k)}=(\beta_{1}^{(k)}\ddd\beta_{p}^{(k)})^{\top}$ be
the $k$-th iterate of $\beta$. The algorithm finds the next update
$\beta^{(k+1)}$ using
\begin{align}
\beta^{(k+1)} & =\beta^{(k)}-\bigg[\frac{\partial Q_{\beta^{(k)}}(\beta^{(k)})}{\partial\beta^{\T}}\bigg]^{-1}Q_{\beta^{(k)}}(\beta^{(k)})\nonumber \\
 & =\beta^{(k)}-\biggl[\frac{\partial\mathbf{U}(\beta^{(k)})}{\partial\beta^{\T}}+\boldsymbol{\Lambda}_{\lambda}(\beta^{(k)})\biggr]^{-1}Q_{\beta^{(k)}}(\beta^{(k)}).\label{eq:newton}
\end{align}
Note that once a component $\beta_{j}^{(k)}$ becomes zero during
the iteration, the term $p_{\lambda}^{\prime}(|\beta_{j}^{(k)}|)/|\beta_{j}^{(k)}|$
in $\boldsymbol{\Lambda}_{\lambda}(\beta^{(k)})$ becomes illy defined.  To continue the iteration, the algorithm would have to stop updating those zero components and simply set their final estimates to zero, and then works only with the nonzero components of $\beta$. This treatment, however, creates a potential problem, that is, once a component of $\beta$ becomes zero, it is permanently deleted and will never again receive updates. To fix this, \citet{hunter2005variable} replaced $(p_{\lambda}^{\prime}(|\tilde{\beta}_{j}|)/|\tilde{\beta}_{j}|)\beta_{j}$
in \eqref{eq:approx} with $(p_{\lambda}^{\prime}(|\tilde{\beta}_{j}|)/(|\tilde{\beta}_{j}|+\epsilon))\beta_{j}$
for some $\epsilon>0$. This leads to a modified $\boldsymbol{\Lambda}_{\lambda}(\widetilde{\beta})=\diag\{p_{\lambda}^{\prime}(|\tilde{\beta}_{1}|)/(|\tilde{\beta}_{1}|+\epsilon)\ddd p_{\lambda}^{\prime}(|\tilde{\beta}_{p}|)/(|\tilde{\beta}_{p}|+\epsilon)\}$
in \eqref{eq:LQA} and \eqref{eq:newton}. 

The algorithms by \citet{fan2001variable} and \citet{hunter2005variable}
suffer from some significant drawbacks: (a) they cannot easily handle
more complex penalty functions, such as the group and sparse group
lassos; (b) the Newton\textendash Raphson update in \eqref{eq:newton}
involves the inversion of the $p\times p$ matrix $\partial\mathbf{U}(\beta^{(k)})/\partial\beta^{\T}+\boldsymbol{\Lambda}_{\lambda}(\beta^{(k)})$, 
which renders the algorithm extremely impractical
for high-dimensional data, when, e.g. $p=100,000$; (c) the update
in \eqref{eq:newton} does not directly produce a sparse solution,
so one needs to manually truncate the $\hat{\beta}_{j}$'s to zero
when $|\hat{\beta}_{j}|<c$ for some threshold $c$, but there is
no theoretical guideline on how to choose the value of $c$, and in
practice it is just set to an arbitrarily small number; and (d) the
convergence properties of the algorithm  in \eqref{eq:newton} were
studied only for the maximum penalized likelihood \citep{hunter2005variable},
but have never been established for regularized estimating equations.

\subsection{Computation for the fixed-point formulation}
\label{sec:connect_REE_FPP_algorithm} 

Suppose $f:\R^{p}\rightarrow\R^{p}$ has Lipschitz constant $L>0$
such that 
\begin{equation}
\|f(\beta)-f(\beta')\|_{2}\leq L\|\beta-\beta'\|_{2},\qquad\text{for all }\beta,\beta'\in\R^{p}.\label{eq:nonexpansive}
\end{equation}
When $L=1$, $f$ is referred to as a nonexpansive mapping and its
set of fixed points $\mathcal{P}=\{\beta:f(\beta)=\beta\}$
is closed and convex ($\mathcal{P}$ can be empty or can contain multiple points; see \cite{ryu2016primer}).
Instead, if $L<1$, $f$ is called a contraction and admits exactly one
fixed point \citep[page 6]{ryu2016primer}. 

A very straightforward algorithm for solving \eqref{eq:fixed_pt}
is the \emph{fixed-point iteration} \citep{picard1890memoire,lindelof1894application,banach1922operations},
also called the \emph{Picard iteration}:
\begin{equation}
\beta^{(k+1)}=f(\beta^{(k)}),\qquad k=0,1,2,\ldots,\label{eq:picard}
\end{equation}
with an initial value $\beta^{(0)}$. One can show that if $f$ is
a contraction with Lipschitz constant $L<1$, the fixed-point iteration
described in Algorithm \ref{alg:Fixed_point_iteration} can converge
to the unique fixed-point $\hat{\beta}$ of $f$ with a geometric
rate \citep[p15,][]{ryu2016primer}:
\[
\|\beta^{(k)}-\hat{\beta}\|\leq L^{k}\|\beta^{(0)}-\hat{\beta}\|.
\]

However, if $f$ is only nonexpansive, the fixed-point iteration \eqref{eq:picard}
may not converge to the set of fixed-points $\mathcal{P}$.
Alternatively, we can use the \emph{Krasnosel'skii\textendash Mann
iteration \citep[KM,][]{mann1953mean,krasnosel1955two}:}
\begin{equation}
\beta^{(k+1)}=(1-\rho)\beta^{(k)}+\rho f(\beta^{(k)}),\qquad k=0,1,2,\ldots,\label{eq:KM}
\end{equation}
with $\rho\in(0,1)$. Assume the set of fixed-points $\mathcal{P}$
is nonempty. Then the KM iteration detailed in Algorithm \ref{alg:KM_iteration}
will yield updates $\beta^{(k)}\rightarrow\hat{\beta}$, for some $\hat{\beta}\in\mathcal{P}$,
that satisfy \emph{Fej\'er monotonicity }
\[
\inf_{\hat{\beta}\in\mathcal{P}}\|\beta^{(k)}-\hat{\beta}\|\rightarrow0.
\]
Moreover, the points yielded by the KM iteration satisfy the fixed-point condition \eqref{eq:fixed_pt} arbitrarily closely, 
\[
\|f(\beta^{(k)}) - \beta^{(k)}\|_{2} \rightarrow 0,
\]
with rate $O(1/k)$. Specifically, we have 
\begin{equation}
\min_{j=0\ddd k}\|f(\beta^{(j)})-\beta^{(j)}\|_{2}^{2}\leq\frac{\|\beta^{(0)}-\hat{\beta}\|}{(k+1)\rho(1-\rho)}.\label{eq:KM_rate}
\end{equation}
Choosing $\rho=1/2$ can maximize $\rho(1-\rho)$, and therefore minimizes
the righthand side of the inequality \eqref{eq:KM_rate}. This suggests
a possible choice $\rho=1/2$, which gives the simple iteration
\[
\beta^{(k+1)}=(1/2)\beta^{(k)}+(1/2)f(\beta^{(k)}),\qquad k=0,1,2,\ldots.
\]

\noindent 
\begin{algorithm}
\SetKwData{Left}{left}\SetKwData{This}{this}\SetKwData{Up}{up}
\SetKwFunction{Union}{Union}\SetKwFunction{FindCompress}{FindCompress}
\SetKwInput{Input}{Input}
\Input{Regularization parameter $\lambda>0$, function $\mathbf{U}$, $\tau > 0$}

Initialize $\beta^{(0)}$;

\For{$k=1,2, \ldots$}{
 $\beta^{(k+1)}=\prox_{\tau\lambda\Omega}(\beta^{(k)}-\tau\mathbf{U}(\beta^{(k)}))$;
}

\caption{fixed-point iteration.\label{alg:Fixed_point_iteration}}
\end{algorithm}

\noindent 
\begin{algorithm}
\SetKwData{Left}{left}\SetKwData{This}{this}\SetKwData{Up}{up}
\SetKwFunction{Union}{Union}\SetKwFunction{FindCompress}{FindCompress}
\SetKwInput{Input}{Input}
\Input{Regularization parameter $\lambda>0$, function $\mathbf{U}$, $\rho\in(0,1)$, $\tau > 0$}

Initialize $\beta^{(0)}$;

\For{$k=1,2, \ldots$}{
 $\beta^{(k+1)}=(1-\rho)\beta^{(k)}+\rho \prox_{\tau\lambda\Omega}(\beta^{(k)}-\tau\mathbf{U}(\beta^{(k)}))$;
}

\caption{Krasnosel'skii\textendash Mann iteration.\label{alg:KM_iteration}}
\end{algorithm}

\subsection{Computation for variational inequality formulation}
\label{sec:REE_to_VIP_algorithm}

We can solve the variational inequality \eqref{eq:VIP} using the
Golden Ratio Algorithm (GRA) proposed by \citet{malitsky2019golden}.
At each iteration, the algorithm only requires the evaluation of $\mathbf{U}$
and $\prox_{\lambda\Omega}$. Algorithm \ref{alg:GRA_fixed} provides
the computational details of this method with a fixed stepsize. 
\noindent 
\begin{algorithm}[h]
\SetKwData{Left}{left}\SetKwData{This}{this}\SetKwData{Up}{up}
\SetKwFunction{Union}{Union}\SetKwFunction{FindCompress}{FindCompress}
\SetKwInput{Input}{Input}
\Input{Lipschitz constant $L$, function $\mathbf{U}$.}

Initialize $\beta^{(1)}$ and $\bar{\beta}^{(0)}$, golden ratio $\phi=\frac{\sqrt{5}+1}{2}$, fixed step size $t\in(0,\frac{\phi}{2L}]$;

\For{$k=1,2, \ldots$}{
Compute $\bar{\beta}^{(k)}=\frac{(\phi-1)\beta^{(k)}+\bar{\beta}^{(k-1)}}{\phi}$;

 $\beta^{(k+1)}=\prox_{t\lambda\Omega}(\bar{\beta}^{(k)}-t\mathbf{U}(\beta^{(k)}))$;
}

\caption{Golden ratio algorithm with a fixed step size.\label{alg:GRA_fixed}}
\end{algorithm}
Followed from Theorem 1 of \citet{malitsky2019golden}, we know that
if $\mathbf{U}$ in \eqref{eq:VIP} is monotone, i.e. 

\[
\langle\mathbf{U}(\beta)-\mathbf{U}(\beta'),\beta-\beta'\rangle\geq0,\qquad\text{for all }\beta,\beta'\in\R^{p},
\]
and is $L$-Lipschitz continuous, i.e., satisfies \eqref{eq:nonexpansive} with $f$ replaced by $\mathbf{U}$,
then with arbitrary initialization $\beta^{(1)}$, $\bar{\beta}^{(0)}\in\R^{p}$
and a fixed stepsize $t\in(0,\frac{\phi}{2L}]$, the sequences $(\beta^{(k)})$
and $(\bar{\beta}^{(k)})$ generated by Algorithm \ref{alg:GRA_fixed}
converge to the solution of \eqref{eq:VIP} with rate $O(1/k)$.

Algorithm \ref{alg:GRA_fixed} employs a fixed stepsize $t\in(0,\frac{\phi}{2L}]$,
which requires the knowledge of the Lipschitz constant $L$. If the
value of $L$ is not available, one can adopt an adaptive stepsize
version of the GRA algorithm for solving \eqref{eq:VIP}; see details
in Algorithm \ref{alg:GRA_adaptive}.  This approach does not require
a line-search. The adaptive GRA computes the stepsizes in each iteration by approximating an inverse local Lipschitz constant of $\mathbf{U}$, and has the same computational cost as the fixed stepsize version.  \citet{malitsky2019golden} showed that, even when $\mathbf{U}$ is
only locally Lipschitz,
with arbitrary initialization $\beta^{(1)}$ and $\bar{\beta}^{(0)}\in\R^{p}$,
  the updating sequences $(\beta^{(k)})$  and $(\bar{\beta}^{(k)})$
generated by Algorithm \ref{alg:GRA_adaptive} can converge to a solution
of \eqref{eq:VIP} with rate $O(1/k)$.

\noindent 
\begin{algorithm}
\SetKwData{Left}{left}\SetKwData{This}{this}\SetKwData{Up}{up}
\SetKwFunction{Union}{Union}\SetKwFunction{FindCompress}{FindCompress}
\SetKwInput{Input}{Input}
\Input{golden ratio $\bar{t}>0$, $\phi=\frac{\sqrt{5}+1}{2}$, $\varphi\in (1,\phi]$, $\rho=\frac{1}{\varphi}+\frac{1}{\varphi^2}$, function $\mathbf{U}$.}

Initialize $\beta^{(0)}$ and $\beta^{(1)}=\bar{\beta}^{(0)}$, stepsize $t_0=\frac{\|\beta^{(1)}-\beta^{(0)}\|}{\|\mathbf{U}(\beta^{(1)})-\mathbf{U}(\beta^{(0)})\|}$, $\theta_0 = 1$;

\For{$k=1,2, \ldots$}{
Find the step size
\[
t_{k}=\min \left\{\rho t_{k-1}, \frac{\varphi \theta_{k-1}}{4 t_{k-1}} \frac{\|{\beta}^{(k)}-{\beta}^{(k-1)}\|^{2}}{\|\mathbf{U}({\beta}^{(k)})-\mathbf{U}({\beta}^{(k-1)})\|^{2}}, \bar{t}\right\}.
\]

Update
\begin{gather*}
\bar{\beta}^{(k)}=\frac{(\varphi-1)\beta^{(k)}+\bar{\beta}^{(k-1)}}{\varphi}, \\
\beta^{(k+1)}=\prox_{t_k \lambda\Omega}(\bar{\beta}^{(k)}-t_k\mathbf{U}(\beta^{(k)})).
\end{gather*}

Update $\theta_{k}=\frac{t_{k}}{t_{k-1}} \varphi$.
}

\caption{Adaptive golden ratio algorithm.\label{alg:GRA_adaptive}}
\end{algorithm}
%


\bibliographystyle{nameyear}
\bibliography{eelasso}

\end{document}

%% file: macros.tex
\global\long\def\argmin{\operatorname*{arg\,min}}%

\global\long\def\argmax{\operatorname*{arg\,max}}%

\global\long\def\sgn{\operatorname*{sgn}}%

\global\long\def\balpha{\boldsymbol{\alpha}}%

\global\long\def\bbeta{\boldsymbol{\beta}}%

\global\long\def\bgamma{\boldsymbol{\gamma}}%

\global\long\def\bdelta{\boldsymbol{\delta}}%

\global\long\def\bzeta{\boldsymbol{\zeta}}%

\global\long\def\bvpi{\boldsymbol{\varpi}}%

\global\long\def\btheta{\boldsymbol{\theta}}%

\global\long\def\bxi{\boldsymbol{\xi}}%

\global\long\def\bveps{\boldsymbol{\varepsilon}}%

\global\long\def\bomega{\boldsymbol{\omega}}%

\global\long\def\bvphi{\boldsymbol{\varphi}}%

\global\long\def\boldeta{\boldsymbol{\eta}}%

\global\long\def\bDelta{\boldsymbol{\Delta}}%

\global\long\def\bPhi{\boldsymbol{\Phi}}%

\global\long\def\bSigma{\boldsymbol{\Sigma}}%

\global\long\def\bTheta{\boldsymbol{\Theta}}%

\global\long\def\bA{\mathbf{A}}%

\global\long\def\bH{\mathbf{H}}%

\global\long\def\bI{\mathbf{I}}%

\global\long\def\bM{\mathbf{M}}%

\global\long\def\bS{\mathbf{S}}%

\global\long\def\bT{\mathbf{T}}%

\global\long\def\bW{\mathbf{W}}%

\global\long\def\bU{\mathbf{U}_{n}}%

\global\long\def\bX{\mathbf{X}}%

\global\long\def\bY{\mathbf{Y}}%

\global\long\def\bZ{\mathbf{Z}}%

\global\long\def\ba{\mathbf{a}}%

\global\long\def\bb{\mathbf{b}}%

\global\long\def\bc{\mathbf{c}}%

\global\long\def\bd{\boldsymbol{d}}%

\global\long\def\bg{\mathbf{g}}%

\global\long\def\br{\mathbf{r}}%

\global\long\def\bs{\mathbf{s}}%

\global\long\def\bu{\mathbf{u}}%

\global\long\def\bv{\mathbf{v}}%

\global\long\def\bw{\mathbf{w}}%

\global\long\def\bx{\mathbf{x}}%

\global\long\def\by{\mathbf{y}}%

\global\long\def\bz{\mathbf{z}}%

\global\long\def\cone{\mathscr{C}}%

\global\long\def\cz{\mathscr{Z}}%

\global\long\def\E{\mathbb{E}}%

\global\long\def\expt{\mathscr{E}}%

\global\long\def\pr{\mathrm{P}}%

\global\long\def\var{\mathrm{var}}%

\global\long\def\real{\mathbb{R}}%

\global\long\def\prob{\mathbb{P}}%

\global\long\def\bbX{\mathbb{X}}%

\global\long\def\T{\top}%

\global\long\def\cmpt{{\scriptscriptstyle \text{c}}}%

\global\long\def\veps{\varepsilon}%

\global\long\def\event{\mathcal{E}}%

\global\long\def\actset{\mathcal{A}}%

\global\long\def\orc{{\scriptstyle \text{o}}}%

\global\long\def\initial{{\scriptstyle \text{initial}}}%

\global\long\def\lasso{{\scriptstyle \text{lasso}}}%

\global\long\def\zero{\mathbf{0}}%

\global\long\def\one{\mathbf{1}}%

\global\long\def\ddd{,\ldots,}%

\global\long\def\bb{\mathbf{b}}%

\global\long\def\bv{\mathbf{v}}%

\global\long\def\bx{\mathbf{x}}%

\global\long\def\bz{\mathbf{z}}%

\global\long\def\bB{\mathbf{B}}%

\global\long\def\bX{\mathbf{X}}%

\global\long\def\bbeta{\boldsymbol{\beta}}%

\global\long\def\bEta{\boldsymbol{\eta}}%

\global\long\def\bbk{\boldsymbol{\beta}^{(k)}}%

\global\long\def\ind{\mathbb{I}}%

\global\long\def\R{\mathbb{R}}%

\global\long\def\S{\mathbb{S}}%

\global\long\def\Sb{\bar{\S}}%

\global\long\def\Sbp{\bar{\S}^{\perp}}%

\global\long\def\Sp{\S^{\perp}}%

\global\long\def\diag{\operatorname{diag}}%

\global\long\def\sign{\operatorname{sign}}%

\global\long\def\prox{\operatorname{prox}}%

\global\long\def\Proj{\operatorname{Proj}}%

\global\long\def\smallo{{\scriptstyle \mathcal{O}}}%

\global\long\def\bigo{\mathcal{O}}%

\global\long\def\dom{\mathrm{dom}}%

\global\long\def\ku{\kappa_{\mathbf{U}}}%

\global\long\def\tu{\tau_{\mathbf{U}}^{2}}%

\global\long\def\bX{\mathbf{X}}%

\global\long\def\bU{\mathbf{U}}%

\global\long\def\bUn{\bU_{n}}%

\global\long\def\Xij{X_{ij}}%

\global\long\def\eij{\varepsilon_{ij}}%

\global\long\def\eji{\varepsilon_{ji}}%

\global\long\def\Yij{Y_{ij}}%

\global\long\def\Yji{Y_{ji}}%

\global\long\def\piij{{\pi(i),\pi(\lfloor n/2\rfloor+i)}}%

\global\long\def\epiij{\varepsilon_{\piij}}%

\global\long\def\RR{\mathbb{R}}%

\global\long\def\barm{\overline{m}}%

\global\long\def\bfz{\mathbf{0}}%

\global\long\def\Mndelta{\mathbb{M}_{n}(\delta)}%

\global\long\def\bXpiij{ X_{\piij}}%

\global\long\def\sumij{\underset{i\neq j}{\sum\sum}}%

\global\long\def\PP{\mathbb{P}}%

\global\long\def\xii{\xi_{i}}%

\global\long\def\calS{\mathcal{S}}%

\global\long\def\bW{\mathbf{W}}%

\global\long\def\calG{\mathcal{G}}%

\global\long\def\bu{\mathbf{u}}%

\global\long\def\bv{\mathbf{v}}%

\global\long\def\bw{\mathbf{w}}%

\global\long\def\bvg{\bv_{g}}%

\global\long\def\calS{\mathcal{S}}%

\global\long\def\calN{\mathcal{N}}%

\global\long\def\calNg{\calN_{g,1/2}}%

\global\long\def\Rn{R_{n}}%

\global\long\def\Dn{D_{n}}%

\global\long\def\elln{l_{n}}%

\global\long\def\MnR{\mathbb{M}(\Rn)}%

\global\long\def\MnpR{\mathbb{M}^{+}(\Rn)}%

\global\long\def\nc{\textup{nc}}